\documentclass[10pt,twocolumn,preprintnumbers,nofootinbib,aps,prd]{revtex4-2}

\usepackage{graphicx}
\usepackage{amsmath}
\usepackage{srcltx}
\usepackage[utf8]{inputenc}
\usepackage[colorlinks,pdfusetitle]{hyperref}
\usepackage{times}
\usepackage{amssymb}
\usepackage{lineno}

\newcommand{\lsim}{\lesssim}
\newcommand{\gsim}{\gtrsim}
\newcommand{\vphi}{\varphi}
\newcommand{\eq}[1]{Eq.~(\ref{#1})}

\newcommand{\ord}[1]{\mathcal{O}{(#1)}}
\newcommand{\beq}{\begin{equation}}
	\newcommand{\eeq}{\end{equation}}

\newcommand{\e}[1]{\times10^{#1}}
\newcommand{\orcid}[1]{\href{https://orcid.org/#1}{#1}}

\begin{document}
	
\title{\boldmath How fast can protons decay?}
	
\author{Hooman Davoudiasl}
\thanks{\orcid{0000-0003-3484-911X}}
\email{hooman@bnl.gov}
	
\author{Peter B.~Denton}
\thanks{\orcid{0000-0002-5209-872X}}
\email{pdenton@bnl.gov}	
	
\affiliation{High Energy Theory Group, Physics Department \\ Brookhaven National Laboratory, Upton, NY 11973, USA}

\begin{abstract}	
Current laboratory bounds imply that protons are extremely long-lived. However, this conclusion may not hold for all time and in all of space. We find that the proton lifetime can be $\sim 15$ orders of magnitude shorter in the relatively recent past on Earth, or at the present time elsewhere in the Milky Way. A number of terrestrial and astrophysical constraints are examined and potential signals are outlined. We also sketch possible models that could lead to spatial or temporal variations in the proton lifetime. A positive signal could be compelling evidence for a new long range force of Nature, with important implications for the limitations of fundamental inferences based solely on laboratory measurements.
\end{abstract}

\maketitle

\section{Introduction}
Longevity of matter is important for having a Universe that is 13.8 billion years old \cite{ParticleDataGroup:2022pth} and has not qualitatively changed for the past few billion years.  This has allowed for steady environments like our solar system, and the emergence of complex life forms within it.  The stability of protons was initially assumed to be a law of Nature and it took decades before this assumption was questioned and became the subject of experimental investigation  \cite{Goldhaber:1981be}.  The early experimental bounds on the proton lifetime, $\gsim 10^{20}$~yr \cite{Reines:1954pg}, established that its decay is at best a remarkably rare process.   The current lower bound on the lifetime of the proton is extremely large, with the most stringently constrained channel given by 
\beq
\tau (p\to\pi^0 e^+) > 2.4 \times 10^{34}~\text{yr}\,,
\label{ptopie}
\eeq
at 90\% confidence level (CL) from the Super-Kamiokande (SK) \cite{Super-Kamiokande:2020wjk} data.  Such lower bounds are generally taken to imply that any mechanism for proton decay is either at very high scales, as in Grand Unified Theories (GUTs) \cite{Georgi:1974sy,Pati:1974yy,Dimopoulos:1981yj}, or else well sequestered from the Standard Model (SM) particles.  

It is noteworthy that the above limit is about 24 orders of magnitude larger than the age of the Universe!  Such an enormous ``margin of safety" makes one wonder what the minimum proton lifetime would have to be, in order to avoid conflict with established data on cosmological evolution and astrophysics.  And, if all other considerations do not require a lifetime near the bound (\ref{ptopie}), could the laboratory measurements be a reflection of {\it local} effects, in the vicinity of the solar system, or peculiar to the current cosmological era?   

The above considerations lead us to ask: How fast could protons have decayed in the past on Earth, or decay today far away from us?  In this paper, we attempt to answer this question.  We examine a number of constraints including geological data, cosmological bounds on decaying relics, and astrophysical observations.  What we find is that protons could have decayed faster, by many orders of magnitude, here on Earth until relatively recently and that this could be happening elsewhere in the Milky Way, and beyond.  

We will provide simple models that show the above variations in the proton lifetime can be realized in a fairly straightforward fashion, if one invokes new ultralight scalars with feeble couplings to dark sectors. Ultralight bosons have rich phenomenology touching on dark matter -- see Ref.~\cite{Antypas:2022asj} for a recent review on ultralight dark matter -- and lead to unique phenomenological signatures such as superradiance, see Ref.~\cite{Brito:2015oca} for a comprehensive review of superradiance.
We also outline some of the expected signals of proton decay due to a coupling to an ultralight scalar.  A detection of such effects could be compelling evidence for ultralight scalars and long range forces that have yet to be discovered.  

\begin{figure*}
\centering
\includegraphics[width=\textwidth]{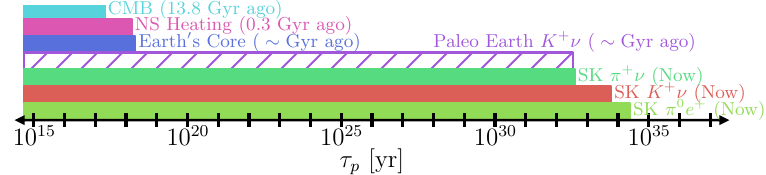}
\caption{The various constraints on the proton lifetime along with the approximate timescale; see the text for details.
Existing constraints are in solid colors and potential future constraints are shown hatched.
Each constraint probes a different region of space and time and are thus complementary.
The NS constraint may also improve in coming years with JWST observations.
}
\label{fig:schematic}
\end{figure*}

\section{The main idea}
We will assume that  proton decay to a dark state \cite{Davoudiasl:2014gfa,Heeck:2020nbq,Fajfer:2020tqf} may be fast, if kinematically allowed. For the purposes of demonstration in this work, we will assume that the dark state is a fermion $\psi$, with no gauge charges, but having baryon number $B(\psi) = + 1$ \cite{Heeck:2020nbq}, which we take to be associated with a good symmetry. No cosmological abundance is assumed for $\psi$, and for parameters relevant to the constraints derived below, we find this assumption is valid  by many orders of magnitude. Hence, proton decay processes such as $p\to \pi^+ \psi$  would be baryon number preserving.  On general grounds, we would then expect that the conventional decay channels containing only SM states, such as $p\to \pi^0 e^+$, to be suppressed relative to the dark decay.  If in the distant past, or far from the Earth, the dark fermion mass satisfies $m_\psi < m_p - m_\pi$, with $m_p \approx 0.94$~GeV and $m_\pi \approx 0.14$~GeV, then proton decay into a $\psi$ final state could be much faster than implied by laboratory experiments.  Other decay channels, like $p\to K^+\psi$, can also be considered with straightforward modifications of the above.  

We will consider two cases where the above  circumstance may be realized, both requiring an ultralight scalar field: {\it (I)} One could entertain the possibility that proton decay was much faster in the distant past, everywhere in the Universe.  {\it (II)} Alternatively, one could have a more parochial view and think of the proton decay rate as controlled by the local Galactic environment around our solar system.  

Later on, we will present models that can in principle realize cases (I) and (II) above.  These models generically also lead to fast neutron decay in nuclei.  Hence, one can consider the more general case of phenomenological constraints on fast nucleon decay, setting current laboratory bounds aside.  Often, the decay rates for the proton and neutron are close and differ by $\ord{1}$ factors.  There may be environments in which neutron decays would constrain these models more tightly, such as in a neutron star (NS) with a sub-dominant proton population; in these scenarios one would also end up with an implied bound on the proton lifetime that may be more stringent than a model-independent one.   However, we will focus on protons in what follows, but the above proviso should be kept in mind when a particular model is considered.  

\section{Novel constraints}
As a point of reference for the constraints derived below, we cite the current 90\% CL laboratory lower bound \cite{Super-Kamiokande:2013rwg,ParticleDataGroup:2022pth}
\beq
\tau(p\to \pi^+ \nu) > 3.9 \times 10^{32}~\text{yr}\,,
\label{ptopinu}
\eeq
which would mimic $p \to \pi^+ \psi$, because both $\nu$ and $\psi$ would go undetected as missing energy.  We will next consider other constraints that can be obtained, independently of the contemporary experimental result (\ref{ptopinu}).  

\subsection{Geology} Let us begin close to home, here on Earth.  Stringent limits on proton decay -- like those in (\ref{ptopie}) and (\ref{ptopinu}) -- have existed only for the past few decades.  One could ask how short the proton lifetime could have been before the modern era.  We focus on the heat output of the Earth and consider how much more heat it would take to melt its solid iron core.  

A recent estimate using a model of heat output and transfer in the Earth suggest that the Earth's inner core has been solid for roughly the past 1-1.3 Gyr \cite{Zhang_2020}.
Other estimates in the literature range from 0.7 to 4.2 Gyr \cite{stacey2007revised,ohta2016experimental,konopkova2016direct}.
Independent magnetic measurements suggest an increase in the Earth's magnetic field 1-1.5 Gyr ago \cite{biggin2015palaeomagnetic}, which is consistent with this picture, although there is also a brief period of inconsistency around 0.6 Gyr ago  \cite{bono2019young}.

The melting point of the iron core, which is under extreme pressure $\sim 330$~GPa , is estimated to be $6230\pm 500$~K; the core temperature is expected to be close to this value  \cite{doi:10.1126/science.1233514}.  Conservatively, we then estimate the amount of extra heat it would take to raise this temperature by $\sim 50\%$, sufficient to melt the inner core.  Assuming a black body radiation approximation, we then need to have about 4 times more heat generated.  The thermal flow out of the core is estimated to be $\sim 10^{13}$~W \cite{Zhang_2020}, which implies that we need an excess of $\sim 4\times 10^{13}$~W.  The core mass is about $10^{26}$~g.  The most abundant iron isotope is $~^{56}_{26}~\text{Fe}$ and hence about $46\%$ of this mass is made up protons for a total number of $N_p^{\rm core} \sim 3 \times 10^{49}$. We thus estimate the heat output required to melt the inner core due to proton decay by 
\beq
f_{\rm SM}m_p N_p^{\rm core} \Gamma_p \sim  4\times 10^{13}~\text{J/s}\,,
\label{core-heat}
\eeq
where $\Gamma_p$ is the needed proton decay rate
and $f_{\rm SM}$ denotes the fraction of the energy in proton decay that goes into SM states (not including neutrinos), which we take to be deposited entirely in the Earth.
We estimate $f_{\rm SM}\sim 0.5$.  (This estimate is fair in the case of our sample decay mode, where $\pi^+$ takes roughly half the energy and stops over distances much smaller than the size of the core \cite{ParticleDataGroup:2022pth}.)
We then find the required proton lifetime $\tau_p \equiv \Gamma_p^{-1}$ for keeping the inner core solid to be given by   
\beq
\tau_p \gsim 2 \times 10^{18}~\text{yr}\,.\quad\quad\text{(Solid Earth Core)}
\label{taup-Earth-core}
\eeq                              
This constraint implies that the proton decay rate cannot have been faster than this for a significant amount of time over the last $\sim$billion years, but does not exclude exotic scenarios where the rate was briefly much higher.

\subsection{Astrophysics}
There are many potential places to look for anomalous heat generation is astrophysical settings.  However, we will focus on neutron star (NS) heating as a promising signal.  In cases where there is sufficient information on the NS galactic environment, observations of the effects assigned to fast proton decay can be interpreted within the context of model classes (I) and (II) above, leading to more insights on the origin of the phenomenon.  

To obtain a constraint, we will use the upper bound $T < 42000$~K on the temperature of the coldest known pulsar PSR J2144–3933 which is $\sim0.3$ Gyr old, based on the observational data form the Hubble Space Telescope \cite{Guillot:2019ugf}.  For this analysis, a mass of $1.4 M_\odot$ was assumed, where $M_\odot \approx 2 \times 10^{33}$~g is the solar mass.  As we are focused on the proton lifetime, we will only consider the protons in the NS which is about $10\%$ of the total baryon content \cite{Bell:2019pyc} across most of the star.  This yields $N_p^{\rm NS} \sim 2 \times 10^{56}$ as the number of protons in the NS.  Assuming the NS is in a steady state, we expect the temperature of the NS to be given by 
\beq
f_{\rm SM} m_p N_p^{\rm NS} \Gamma_p \approx 4 \pi R_{\rm NS}^2\sigma_{\rm SB} T^4\,,
\label{steadystate}
\eeq
where $\sigma_{\rm SB} = \pi^2/60$ is the Stefan-Boltzmann constant and $R_{\rm NS}=11$~km is the assumed radius of the NS \cite{Guillot:2019ugf}.  We thus find
\beq
\tau_p \gsim 1.5\times 10^{18}~\text{yr}\,. \quad\quad \text{(NS Heating)}
\label{taup-NS-heating}
\eeq 
Thus, the proton decay rate could not have been much faster than allowed by the geological constraint (\ref{taup-Earth-core}), over the last few hundred million years as this NS, which is near the Earth, orbited the Milky Way.  Future observations, for example by the James Webb Space Telescope (JWST), have been projected potentially to be able to measure NS temperatures about an order of  magnitude lower \cite{Chatterjee:2022dhp,Raj:2024kjq}. This could lead to an $\ord{10^4}$ improvement of the bound (\ref{taup-NS-heating}).

\subsection{Cosmology}
We also look to cosmology to constrain proton decay.  To this end, we examine existing lower bounds on DM lifetimes.  The constraints may rely on diffuse or line emission of radiation from the decay products of DM \cite{Bell:2010fk} or employ constraints from Cosmic Microwave Background (CMB) \cite{Slatyer:2016qyl,Xu:2024vdn}.  The CMB bounds are generally stronger and also avoid issues regarding the spatial distribution of the input data.  To use the DM bounds, we note that the proton can be treated as a $\sim$~GeV scale relic that carries $\Omega_b/\Omega_{DM}\sim 0.2$ fraction of the DM abundance.  Using the results of Ref.~\cite{Xu:2024vdn}, we then roughly estimate 
\beq
\tau_p \gsim 2\times 10^{17}~\text{yr}\,, \quad\quad \text{(CMB)}
\label{taup-CMB}
\eeq  
where we have used the constraint for $\sim 100$~MeV $e^+e^-$ decay final states.  This is not an exact constraint for the case of proton decay, given that its final states do not necessarily include $e^+e^-$.  However, up to $\ord{1}$ factors, this gives a reasonable estimate of the CMB bound on proton decay in high redshift $z\gtrsim100$ environments.

We briefly mention that Big Bang Nucleosynthesis (BBN) is also sensitive to fast proton decay at times even before the CMB. 
Due to lower precision measurements and shorter timescales in play, however, the constraints are many orders of magnitude weaker than those of the CMB.
In addition, observations of Pop-II stars in the Milky Way, which formed very early in the Universe ($\lsim 1$~Gyr after the Big Bang) and have low metalicity \cite{Edvardsson:1994tj,2015A&A...584A..86S}, can constrain the proton lifetime at the $\tau_p\sim10^{15}$ yr level.  This is due to the inferred abundance of boron in very old stars \cite{Edvardsson:1994tj}, which can be enhanced by the decay of carbon, as one of its protons decays.  The above bound is less competitive than other constraints in the Milky Way, but does probe the  intermediate ages of the Universe, before the geological time scales considered here.

\subsection{Paleo detectors} Assuming a 10\% systematic uncertainty on the kaon production rate from cosmic rays, we find that the results of Ref.~\cite{Baum:2024sst} imply that scanning 100 g of Gyr olivine from the Earth could yield a limit on the proton lifetime of
\begin{equation}
\tau_p\gtrsim 3\e{32}~\text{yr}\,,\quad\quad\text{(Paleo Earth,   Projection)}
\end{equation}
assuming a background rate of 400 kaons/100 g/Gyr and a 10\% systematic uncertainty on the background.
While such a study does not yet exist, it is feasible to be done in the coming years and would represent a very significant improvement in the lifetime of the proton on Earth as our solar system has undergone several orbits of the Milky Way over this time, with a period of $\sim 230$~Myr.
Further improvement to surpass SK's limit may also be possible by collecting samples beneath the lunar surface which would further reduce backgrounds from cosmic ray showers.

Such a paleo measurement would provide a constraint on the \emph{total} number of protons decayed over the last Gyr and is thus slightly different from the Earth's core constraint which in principle allows brief periods of rapid proton decay.

Another terrestrial constraint comes from the extremely low ratio of $^{14}$C/$^{12}$C $\sim 2\times 10^{-18}$ abundance  in natural gas measured by BOREXINO \cite{Borexino:1998eqi}.
Proton decay would feed an additional source of $^{14}$C, which itself has a half life of 5700 years, from $^{15}$N.
The $^{15}$N abundance in the Earth is within $1\%$ of that in the atmosphere and is $4\e{-3}$ \cite{10.1093/nsr/nwae201}.
The carbon to nitrogen ratio in the Earth varies \cite{MARTY201256}, but we conservatively take it to be $\sim1$.
Combined, we find that this places a limit on the lifetime of a proton at the level of $10^{19}$ years, but is only applicable over timescales of the last few thousand years.  That is, if fast proton decay stopped much earlier than that, as may be expected in typical underlying models, the produced $^{14}$C abundance would have been efficiently depleted below the observed level.

There are additional means of probing anomalously fast proton decay such as looking at decay products from Andromeda or dwarf spheroidals \cite{Baring:2015sza,Arguelles:2022nbl}; we find that these constraints are not competitive with those listed here, but could conceivably be relevant in alternative implementations of our idea.

\section{Two models}
The underlying assumption in our work -- that protons could have much shorter lifetimes over astronomical time and distance scales -- would not present a compelling possibility unless it can be realized by  feasible physical mechanisms. Below, we will provide sample models that can potentially realize the general scenarios of types (I) and (II), which were briefly described before.  

\subsection{A type (I) model}
This scenario could, for example, be effected by the time evolution of a modulus $\vphi$ with a tiny mass $m_\vphi$ that governs the size of $m_\psi$.  As $\vphi$ evolves, we require that $m_\psi(\vphi)$ be much smaller in the past, billions of years ago, allowing fast proton decay into $\psi$, but sufficiently large today $m_\psi(\vphi)\gtrsim m_p$, so that such decays are shut off.  This typically implies $m_\vphi \gsim H_0$, where $H_0\approx 1.5 \times 10^{-33}$~eV~$\approx (14.5~\text{Gyr})^{-1}$ is the Hubble expansion rate today.    
We assume that the potential for the ultralight scalar $\vphi$ is given by 
\beq
V_\vphi = \frac{1}{2} m_\vphi^2 (\vphi-\vphi_*)^2\,,
\label{Vphi}
\eeq
where $\vphi_*$ is the minimum of the potential.  The scalar starts its oscillation when $m_\vphi \sim 3 H$.  The current cosmological era is marked by accelerated expansion and the Hubble rate $H_0$, which we assume to be due to a cosmological constant.  Hence, if $m_\vphi > 3 H_0$ then $\vphi$ starts to track its potential in the matter dominated era, eventually oscillating around $\vphi_*$ with a period $\Delta t = 2 \pi/m_\vphi$. 

We will assume that there is a Yukawa coupling between $\vphi$ and $\psi$ given by 
\beq
\lambda \,\vphi\, \bar \psi \psi\,,
\label{Yukawa}
\eeq
with $\lambda\ll 1$.  For a small initial value $\vphi_i\ll \vphi_*$ the dark fermion $\psi$ could be light, but as the scalar approaches the minimum of its potential $\psi$ can start to get heavy enough that fast $p \to \pi \psi$ is shut off, assuming that other dimension-6 operators that could mediate its decay are suppressed by very high scales, as is conventionally assumed.  

Let us consider a situation where $\vphi$ could nearly complete a 1/4 period, starting from the onset of dark energy domination at a time $t\sim 2/(3 H_0) \sim 10$~Gyr.  We then demand that $\Delta t \sim \pi/(2 m_\vphi) \sim 4$~Gyr, given that the age of the Universe is about 13.8~Gyr \cite{ParticleDataGroup:2022pth}.  This yields $m_\vphi \sim 6 H_0 \sim 10^{-32}~$eV.  In order to make sure that we do not perturb standard cosmology significantly, we will take the contribution of $V_\vphi$ to dark energy to be small.  A cosmological constant would have energy density $\rho_\Lambda \approx (2\times 10^{-3})^4$~eV$^4$.  Hence, we require 
\beq
\frac{1}{2}m_\vphi^2 \vphi_*^2 \ll \rho_\Lambda\,,
\label{small-Vphi}
\eeq   
which implies 
\beq
\vphi_* \ll 10^{18}~\text{GeV}.
\label{phimax}
\eeq
This then allows for a monotonically increasing $m_\psi$ which would then shut off the decay of protons to $\psi$ at some critical point in the past.
By further modifying the oscillation timescales in the above description, more exotic scenarios can be achieved where proton decay may have been temporarily viable due to the oscillatory nature of the field.

\subsection{A type (II) model}
As an example, take an ultralight scalar $\phi$, with a Compton wavelength $\sim $~kpc, that has a feeble coupling to dark matter (DM).  Then,  $\phi$ would be sourced as a background field whose value depends on DM density over galactic scales.  As in case (I) above, if $m_\psi (\phi) < m_p-m_\pi$ then $p \to \pi^+ \psi$ could have been fast in the past.  This requires the distribution of DM near the solar system change over cosmological times, as generally expected due to local variations -- see {\it e.g.} Ref.~\cite{Lim:2023lss} -- with the present value of $m_\psi(\phi)$ large enough to eliminate the fast proton decay channel.

Here, we adapt a model proposed in Ref.~\cite{Davoudiasl:2018hjw} for sourcing  neutrino masses from the Galactic DM population, but we apply it to generating a spatially varying mass for $\psi$ sourced by the DM, $X$, via a long-range mediator, $\phi$. This can be straightforwardly done by introducing an interaction  
\beq
y\, \phi\, \bar \psi \psi\,,
\label{phipsipsi}
\eeq
where $y\ll 1$ is the Yukawa coupling between the ultralight scalar $\phi$ of mass $m_\phi$ and the dark fermion $\psi$.  As in the model of Ref.~\cite{Davoudiasl:2018hjw}, we also assume that the $\phi$ couples to the DM, $X$, through a feeble interaction 
\beq
g_X \,\phi\, \bar X X\,,  
\label{phiXX}
\eeq
with $g_X\ll 1$.  
We will assume that $m_\phi\sim 10^{-26}$~eV, corresponding to an $\ord{\rm kpc}$ Compton wavelength, encompassing galactic distance scales. One can show that the induced $\phi$ background is given by 
\beq
\phi \sim \frac{g_X n_X}{m_\phi^2}\,,
\label{phi}
\eeq
where $n_X$ is the DM number density.  Substituting the above in \eq{phipsipsi}, we 
may then deduce \cite{Davoudiasl:2018hjw} 
\begin{align}
m_\psi \sim{}& \text{GeV} \left(\frac{y}{10^{-9}}\right) \left(\frac{g_X/m_X}{10^{-19}~\text{GeV}^{-1}}\right) \nonumber \\ 
&\times\left(\frac{\rho_X}{0.4~\text{GeV}\text{cm}^{-3}}\right) \left(\frac{10^{-26}~\text{eV}}{m_\phi}\right)^2 \,, 
\label{mpsi}
\end{align}
where $\rho_X \approx 0.4 ~ \text{GeV}\text{cm}^{-3}$ is the DM density in our Galactic neighborhood \cite{ParticleDataGroup:2022pth,Salucci:2010qr}. 

The above implies that in the outer regions of the Milky Way, or in the intergalactic space, where 
the DM density is much smaller than near the Earth, $m_\psi$ can be much smaller and allow fast proton decay.  Similarly, if the solar system has gone through underdense regions of the Milky Way DM halo,  cosmological epochs marked by fast proton decay could have existed on Earth, in the past.  For example, if such a region has a size of $\ord{\rm kpc}$ -- which may be typical of galactic substructures \cite{Moore:1999nt} -- given that the solar system moves at a speed of $\sim 10^{-3}$ through the Milky Way, we may expect such an epoch to last for a few million years.   

\section{Discussion}
We summarize our results in Fig.~\ref{fig:schematic} which shows the lifetimes disfavored due to different observations.  As we can see, the derived bounds based on general considerations from the CMB, astrophysics, and geology yield limits that are at the level of $\sim 10^{18}$~yr, or less.  A projection based on the ``paleo" technique of looking for particle tracks in terrestrial minerals \cite{Baum:2024sst} could potentially provide a very impressive reach, not far from contemporary laboratory bounds, for stability of protons over cosmological time scales of $\ord{\rm Gyr}$.  However, this is only a simple  projection and requires further investigation, using actual geological samples and data. 

\begin{figure}
\includegraphics[width=\columnwidth]{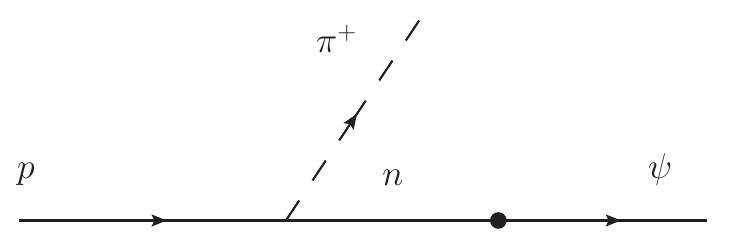}
\caption{An example of a proton decay process, $p\to \pi^+ \psi$,  mediated by the dimension-6 operator (\ref{dim6})  containing a dark fermion $\psi$.  The dot represents    nucleon non-conservation through $n$-$\psi$ mixing.}
\label{fig:pdecay}
\end{figure}

Setting aside the paleo projection, we may ask: What is the ultraviolet scale implied for a minimum proton lifetime lower bound obtained here?  To this end, we need to assume a particular set of higher dimension operators that would mediate our sample decay process $p \to \pi^+ \psi$.  For simplicity, we may choose the  following dimension-6 operator 
\beq
\frac{udd \psi_R^c}{M^2}\,,
\label{dim6}
\eeq
where $u$ and $d$ denote SU(2) singlet up and down quarks, respectively, and $R$ denotes  right-handed chirality.  This operator can lead to the neutron $n$ mixing with $\psi$.  Such a mixing,  through a baryon number preserving low energy hadronic coupling $\pi n p$, can lead to our proton decay process.  This process is schematically illustrated in Fig.~\ref{fig:pdecay}.  

As a phenomenological possibility, it could be tempting to identify the state $\psi_R$ as  a massive Majorana ``right-handed" neutrino \cite{Davoudiasl:2014gfa,Helo:2018bgb}.  In that case, the modulation of  $m_\psi$ may be correlated with changes in the SM neutrino masses across space or time, through a seesaw mechanism.  This could lead to a scenario with  interesting implications, but we do not investigate it further here.    

We may use the results of Ref.~\cite{Davoudiasl:2014gfa} for the interaction (\ref{dim6}),  derived using a chiral perturbation formalism \cite{Claudson:1981gh}, to estimate the size of $M$ for a given proton lifetime set by $p\to \pi^+ \psi$.  If $\psi$ is relatively light, we may expect a lifetime of $\sim 6\times 10^{31}$~yr for $M=10^{15}$~GeV \cite{Davoudiasl:2014gfa}, with the lifetime scaling as $M^4$.  Assuming a lifetime of $\sim 10^{20}$~yr, not far above the bounds derived here, we would then find $M\sim 10^{12}$~GeV, which is far below the typical GUT scales $\gsim 10^{15}$~GeV.
We also note that in a GUT model, a time or spatial variation in the proton decay could also correlate the time variation of other fundamental parameters such as $\alpha$, see e.g.~\cite{Calmet:2001nu,Langacker:2001td,Dent:2001ga,Olive:2002tz,Wetterich:2002ic,Dine:2002ir,Calmet:2002ja,Flambaum:2002de,Fritzsch:2003wf}.

We have not assumed any other interactions for $\psi$ with the SM.  However, it could possibly have couplings that do not lead to nucleon decay. 
 Given that we would typically expect a light $\psi$ in our scenario, one could possibly entertain the idea that this state may be accessible in low energy experiments that probe the GeV scale.  However, we do not elaborate further on this possibility here.   

Finally, we mention that it is conceivable to have a positive signal for time or spatial dependent proton decay.
One approach is a correlated set of measurements such as a geological period of anomalous heating of the Earth's core combined with paleo measurements from different epochs that show no proton decay up to a certain period and do show proton decay over the same period as the Earth's core heating at a consistent rate.
Another approach is if anomalous neutron star heating is observed that is only present in neutron stars in low dark matter densities which is a trend that goes in the opposite direction as typical dark matter heating models.
Other possible combinations of data sets may also lead to convincing positive signals.

\section{Summary}
In this paper, we have taken the point of view that the proton decay rate may vary throughout space and time, due to its dependence on an ultralight modulus.  This modulus could possibly be undergoing cosmological evolution, resulting in a time varying effect.  However, this field may also be sourced, for example by dark matter, in which case one could generally expect both spatial and temporal modulations of the proton decay rate.  In this type of scenario, the lower bound on the proton lifetime deduced from modern terrestrial data  could be a transient effect, and that protons could be decaying much  faster far from the Earth or at much earlier  times.  We then inquired what  lower bounds on the proton lifetime could be derived, independently of  laboratory constraints. 

 We considered constraints from geology, astrophysics, and cosmology,  and found that  geological data provide our strongest constraints for the stability of terrestrial protons over the last billion years.  We expect that ``paleo"    techniques --  based on examination of ancient minerals for signs of charged particle tracks -- could in principle strengthen the local constraints by many orders of magnitude. 
 Examples of models that can realize our  scenarios were also provided.  The relative simplicity of these models make the underlying assumptions of our work worth further theoretical and experimental investigations.  A positive signal for the phenomena outlined here could imply that laboratory experiments  only offer a snapshot of local   microphysics that is susceptible to change, due to the influence of as yet unknown long range forces.  

\vskip0.5cm
\begin{acknowledgments}
We thank Patrick Stengel for helpful comments.
This work is supported by the US Department of Energy under Grant Contract DE-SC0012704. 
No digital data were generated for the results presented in this work.
\end{acknowledgments}

\bibliography{fast}

\end{document}